\documentstyle[aps,twocolumn,psfig]{revtex}

\begin{document}
\draft

\title{Symmetry and dimension of the magnon dispersion
of inorganic spin-Peierls systems}

\author{G\"otz S.~Uhrig}
\address{Laboratoire de Physique des Solides, Universit\'e Paris-Sud,
 b\^at.~510, F-91405 Orsay}
\address{present address: 
Institut f\"ur Theoretische Physik,
Universit\"at zu K\"oln, D-50937 K\"oln}

\date{\today}

\maketitle

\begin{abstract}
The  data on the dispersion of the 
magnetic excitations of CuGeO$_3$ in the spin-Peierls D-phase
are analyzed. On the basis of the lattice structure it is shown
that even along the chains the $d=2$ character cannot be neglected.
The symmetry of the dispersion differs from the one assumed so far.
The magnetic resonance data is reinterpreted. The possibility of 
interchain rather than intrachain frustration is discussed.
\end{abstract}

The substance CuGeO$_3$ is the first inorganic example of a spin-Peierls
system \cite{hase93a}. A second compound which is intensively
investigated is $\alpha'$-NaV$_2$O$_5$ \cite{isobe96,fujii96}
An inorganic sample is particularly interesting since 
it allows to employ all the experimental tools
 (for a review, see \cite{bouch96}).
Thus the investigation of CuGeO$_3$ has become
a very active field both in theory and in experiment.

A spin-Peierls system is an antiferromagnetic spin
system coupled to phonons such that below a certain temperature
$T_{\rm\scriptstyle SP}$ the translational
 symmetry is spontaneously broken. The system
becomes dimerized due to a static lattice distortion which renders the
spin coupling alternating along a certain axis. There are stronger bonds
($J$) and weaker bonds ($\lambda J$, $\lambda<1$).
The phenomenon is essentially driven by $d=1$ physics due to
the extreme sensitivity
of a $d=1$ system towards a $2k_{\rm\scriptstyle F}$ perturbation.
 This is manifest
in the superquadratic growth of the energy gain 
$\Delta E \propto -\delta^{4/3}$ with
the dimerization $\delta := (1-\lambda)/(1+\lambda)$ \cite{cross79}.

It is the aim of the present work to show that CuGeO$_3$ 
is only approximately a $d=1$ compound.
 On the basis
of the dispersion perpendicular to the chains this has been
noted by several authors 
(e.g. \cite{nishi94,nishi95,regna95b,regna96a,bouch96}).
What has passed unnoticed so far is that
 the spin-spin couplings perpendicular to the chains, which are formed
along the $c$-axis, influences the dispersion {\em along} the chains
 in a non-trivial, non-constant way.

So far, the deviations of the behavior of CuGeO$_3$ from the physics
of simple nearest-neighbor (NN) coupled Heisenberg chains plus phonons
were attributed to an intrachain frustration, i.e.\ an additional
next-nearest-neighbor (NNN), antiferromagnetic coupling $J_2$ 
\cite{loren94,riera95,casti95,haas95,regna96a}.
But no consensus has been reached which value the relative frustration
$\alpha=J_2/J$ \cite{note1} applies for CuGeO$_3$. The values range from
$\alpha=0.36$ \cite{riera95} to $\alpha=0.17$ \cite{regna96a}
and the values for $J$ from 10.4 meV \cite{nishi94} to 21.5 meV \cite{kuroe96}.
This poses a problem since the NNN coupling represents a marginal 
perturbation which engenders above the critical value $\alpha_c=0.241$
\cite{okamo92} a gap in the spin system
itself without alternating coupling constants.
The high temperature data for the magnetic susceptiblity $\chi(T)$ 
\cite{hase93a,hori94} suggests values above $\alpha_c$
\cite{riera95,fabri97a} whereas there is no indication for a spin gap without
coupling alternation \cite{casti95}. An analysis of the dispersion
$\omega(q_c)$ at small $q_c$ suggests $\alpha=0.17<\alpha_c$ \cite{regna96a}.

The dispersion in b-direction is roughly $0.3$ of the one in c-direction;
the dispersion in a-direction is roughly 1/50. First, we
focus on the situation in the bc-plane. Up to now, the complete dispersions
$\omega(\vec{q})$ is interpreted, at least qualitatively, as being
a sum of two independent contributions $\omega(\vec{q})\approx \omega_b(q_b)
+\omega_c(q_c)$ leading to the schemes in \cite{loosd96a} (inset fig.~2)
and in \cite{bouch96}(fig.~9). 

On the other hand, elastic neutron scattering \cite{hirot94,brade96a}
provides evidence that the dimerization pattern is shifted by one spacing
from one chain to the other, i.e.\ adjacent copper ions move in opposite
directions \cite{hirot94}. A segment of the resulting dimer distribution
is shown in fig.\ 1.
\begin{figure}
\setlength{\unitlength}{1cm}
\begin{picture}(8.2,3.6)(-0.5,2.6)
{\psfig{file=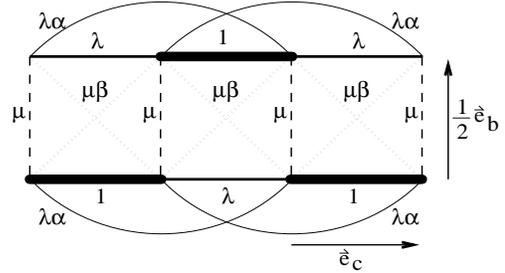,height=6.3cm,width=6.5cm,angle=270}}
\end{picture}
\caption{Segment of the dimerized lattice in the bc-plane. The largest
coupling is set to unity and the weaker couplings are parametrized
as indicated.}
\end{figure}
The bold lines represent the dimers
($J$ is set to unity); the other
lines (thin, dashed, dotted) represent the residual, weaker couplings
($\lambda, \mu < 1$). The line $\lambda$ is the weak bond in the dimerized
chain, $\mu$ is the interchain coupling, $\mu\beta$
is a possible NNN hopping, and $\lambda\alpha$ is the
intrachain frustration discussed before. The microscopic superexchange
paths for $\mu$ and $\mu\beta$ are Cu-O2-O2-Cu in b-direction, see 
\cite{brade96a} fig.~1.  Since from one of the O2 two Cu with different
c-coordinate can be reached, a direct super-exchange ($\mu$) or a 
super-exchange with
a shift along the c-axis ($\mu\beta$) is possible \cite{brade96a}.
One expects that $\beta\approx 0.5$ because there are
two paths for the direct super-exchange and one for each shifted 
super-exchange. 

The analysis starts from the limit of strong dimerization.
The experimental systems are closer to $\delta=0$ than to 
$\delta=1$. But the $\delta=1$ case is a good starting
point due to its simplicity which allows a systematic 
perturbative approach, in principle possible down to arbitrarily
small $\delta>0$. In practice a third order calculation yields
reasonable results down to $\delta\approx 0.1$ 
(see fig.\ 1 in \cite{uhrig96b}).

At $\delta=1$, the ground state
is a product of local singlets on the dimers and a magnon is a 
degenerate local single triplet.
These triplets acquire a dispersion due to the residual couplings
which lift the degeneracy.
To first order one finds a hopping from dimer to dimer
 along the chains of $t_{10}=-\lambda(1-2\alpha)/4$ \cite{uhrig96b}
 and of $t_{11}=-\mu(1-2\beta)/4$ to the nearest dimers in
b-direction. To obtain the corresponding dispersion
one may view the dimers as sites of the effective lattice
 in fig.~2.
\begin{figure}
\setlength{\unitlength}{1cm}
\begin{picture}(8.2,4.5)(0.5,6)
{\psfig{file=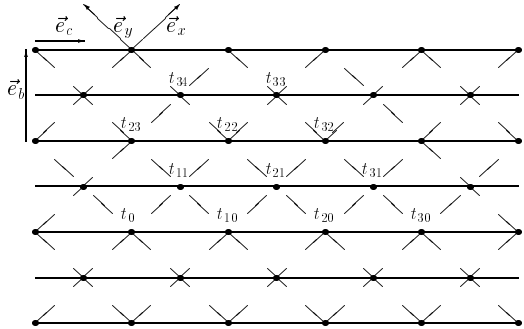,height=12cm,width=9cm}}
\end{picture}
\caption{Effective lattice if the dimers are taken as elementary sites.
The hopping elements refer to hops starting at the site $t_0$.}
\end{figure}
Dashed lines correspond to the processes
$\propto \mu$ and solid lines to processes $\propto \lambda$. 
Starting from $t_0$ one reaches the sites
$t_{10}$, $t_{11}$ and their symmetry equivalents (reflection at 
the c-axis and the b-axis) with the amplitudes given above.
A natural unit cell is spanned by $\vec{e}_x$
and $\vec{e}_y$ as indicated. The dashed lines form a square lattice to
which the solid lines add a component along one of the diagonals. The
dispersion reads $\omega(\vec{q}) = 2t_{11}(\cos(q_x) + \cos(q_y))
+2t_{10}cos(q_x-q_y)$ where $q_{x/y}$ are the components along the 
dual basis vectors belonging to $\vec{e}_x$ and to $\vec{e}_y$,
respectively. Changing to the basis dual to $\{\vec{e}_b, \vec{e}_c\}$
by $q_{x/y} = q_b/2 \pm q_c$ one obtains
\begin{equation}
\label{basic}
\frac{\omega(\vec{q})}{J} = 1 + 2t_{10}\cos(2q_c) + 4t_{11}\cos(q_b/2)
\cos(q_c) \ .
\end{equation}

Eq.~(\ref{basic}) illustrates already the main points of this paper.  A
dispersion of type (\ref{basic}) has minima at $\vec{q}=(0,0,0)$ and
at $(0,1,1/2)$. Saddle points are at $(0,1,0)$ and at $(0,0,1/2)$.
These features stay valid in higher perturbative orders as well. They
change the picture used so far (cf.\ fig.\ 9 in \cite{bouch96})
where the symmetry relations $(0,0,0) \leftrightarrow (0,0,1/2)$
and $(0,1,0) \leftrightarrow (0,1,1/2)$ held.
The fact that the system is actually 2d (or 3d, see below) does not
only lower the dispersion along c by some constant for constant $q_b$
but adds another $q_c$ dependence.  The symmetry about $q_c=\pi$,
which one has in $d=1$ due to dimerization
\cite{uhrig96b}, is broken by the term proportional to $\cos(q_c)$.
Since $t_{11}$ is $\approx 1/7$ of $t_{10}$ the effect is not very large, but
visible.  Physically this mixture of motion in the b- and
c-direction is understandable in fig.~1.  Any hop along b
implies a shift also along c because the dimers alternate.
In addition, the motion along b lowers the observable gap.
Hence a given gap corresponds to a larger dimerization than in $d=1$.

For a more quantitative comparison of theory and experiment we extend
 (\ref{basic}) to third order by standard
degenerate perturbation theory
\begin{mathletters}
\label{result}
\begin{eqnarray}
  t_0 &=& 1 - (4-3\bar\alpha^2)\lambda^2/16 - (4-3\bar\beta^2)\mu^2/8
  - (8-8\bar\alpha \nonumber\\ &-&
  6\bar\alpha^2+3\bar\alpha^3)\lambda^3/64 -
  (8-8\bar\beta-6\bar\beta^2+3\bar\beta^3)\mu^3/32 \nonumber \\ &+&
  (8\bar\alpha-16\bar\beta +18\bar\alpha\bar\beta+9\bar\beta^2-12
  \bar\alpha\bar\beta^2)\lambda\mu^2/32 \nonumber \\ &+&
  \lambda^4/128 - 3\lambda^2\mu^2/16 - 3\mu^4/64 \\ t_{10} &=&
  -\bar\alpha\lambda/4 - \lambda^2/8 -\bar\beta^2 \mu^2/16
  -(4-4\bar\alpha-\bar\alpha^3)\lambda^3/64 \nonumber \\ &-&
  (2-\bar\beta)\bar\beta^2\mu^3/32 +(6\bar\alpha-6\bar\beta
  +\bar\beta^2+ 2\bar\alpha\bar\beta^2)\lambda\mu^2/32 \nonumber \\ 
  &+& \lambda^4/128 + \lambda^2\mu^2/32 + 3\mu^4/128 \\ t_{11} &=&
  -\bar\beta\mu/4 -\bar\alpha\bar\beta\lambda\mu/16 -\mu^2/8
  +(4\bar\alpha + 12\bar\beta -4\bar\alpha\bar\beta \nonumber \\ &+&
  7\bar\alpha^2\bar\beta) \lambda^2\mu/128
  +(\bar\alpha-\bar\beta)\lambda\mu^2/8 - (8-16\bar\beta \nonumber \\ 
  &-& 7\bar\beta^3)\mu^3/128 +5\lambda^2\mu^2/128 +3\mu^4/128 \\ 
  t_{20} &=& -\bar\alpha^2\lambda^2/32 -
  (2-\bar\alpha)\bar\alpha^2\lambda^3/64 \nonumber \\ 
  &-&3\bar\alpha\bar\beta^2\lambda\mu^2/64 +3\lambda^4/256 \\ t_{21}
  &=& -\bar\alpha\bar\beta\lambda\mu/16 -(8\bar\alpha - 8\bar\beta
  +8\bar\alpha\bar\beta +5\bar\alpha^2\bar\beta)/\lambda^2\mu/256
  \nonumber \\ 
  &+&(2\bar\alpha-2\bar\beta-2\bar\alpha\bar\beta+\bar\alpha\bar\beta^2)
  \lambda\mu^2/64 -3\bar\beta^3\mu^3/256 \nonumber \\ &+&
  3\lambda^2\mu^2/128 \\ t_{22} &=& -\bar\beta^2\mu^2/32 -
  3\bar\alpha\bar\beta^2\lambda\mu^2/64
  -(2-\bar\beta)\bar\beta^2\mu^3/64 \nonumber \\ &+&3\mu^4/256 \\ 
  t_{23} &=& -\bar\beta^2\mu^2/16 -
  3\bar\alpha\bar\beta^2\lambda\mu^2/64
  +(4-2\bar\beta+\bar\beta^2)\bar\beta\mu^3/32 \nonumber \\ 
  &+&3\mu^4/128 \\ t_{30} &=& -\bar\alpha^3\lambda^3/128 \qquad t_{31}
  \, =\, -3\bar\alpha^2\bar\beta\lambda^2\mu/128 \\ t_{32} &=&
  -3\bar\alpha\bar\beta^2\lambda\mu^2/128 \qquad t_{33} \, =\,
  -\bar\beta^3\mu^3/128 \\ t_{34} &=& -3\bar\beta^3\mu^3/128\ .
\end{eqnarray}
\end{mathletters}
All frustration terms have been included with $\bar\alpha = 1-2\alpha$
and $\bar\beta = 1-2\beta$. This is not too difficult once one has the
perturbation terms for $\alpha=0,\beta=0$ because
the frustration terms link the same dimers and they have the same
effects on them except for the matrix elements. The
dispersion reads
\begin{eqnarray}
\label{dispers}
\frac{\omega(\vec{q})}{J} &=& [ t_0 + 2t_{10}\cos(2q_c)
+2t_{20}\cos(4q_c) + 2t_{30}\cos(6q_c) \nonumber \\ &+&\!
4\cos(q_b/2)( t_{11}\!\cos(q_c) \!+\! t_{21}\!\cos(3q_c) \! +\!
t_{31}\!\cos(5q_c)) \nonumber \\ &+& 2\cos(q_b)(t_{23}
+2t_{22}\cos(2q_c) +2t_{32}\cos(4q_c)) \nonumber \\ &+& 4\cos(3q_b/2)(
t_{34}\cos(q_c) + t_{33}\cos(3q_c))] . \ 
\end{eqnarray}

For $\mu=0, \alpha=0$, (\ref{result}), (\ref{dispers}) can be checked
with the result for a dimerized NN chain \cite{harri73}.  For
$\mu=0,\lambda\alpha = J/J_\perp, \lambda=0$ one retrieves the result
for a NN spin ladder \cite{reigr94}.  The upper chain is made up by
the first sites of the dimers; the lower chain of the second sites.
For $\mu=0,\lambda\alpha = J/J_\perp$ and arbitrary $\lambda$, a general
zigzag chain is described. This should give an excellent description
of the magnon dispersion in Cu$_2$(C$_5$H$_{12}$N$_2$)$_2$Cl$_4$
\cite{chabo96}.

Another special case is (($\alpha=1/2$ or $\lambda=0$) and
($\beta=1/2$ or $\mu=0$)) where brackets, `ands', and `ors' are
logical signs. The singlet product state is the ground state in the
$d=1$ case $\mu=0, \lambda\le1$ \cite{majum69c}.  The same can be
proven for the $d=2$ lattice in fig.~2 if $\alpha=\beta=1/2$.  For
this case the fourth order terms are included in (\ref{result}).
  It is reasonable to
include the fourth order terms if $\alpha, \beta$ are close to $1/2$.

\begin{figure}
  \setlength{\unitlength}{1cm}
\begin{picture}(8.2,5.4)(-1.2,0.4)
  {\psfig{file=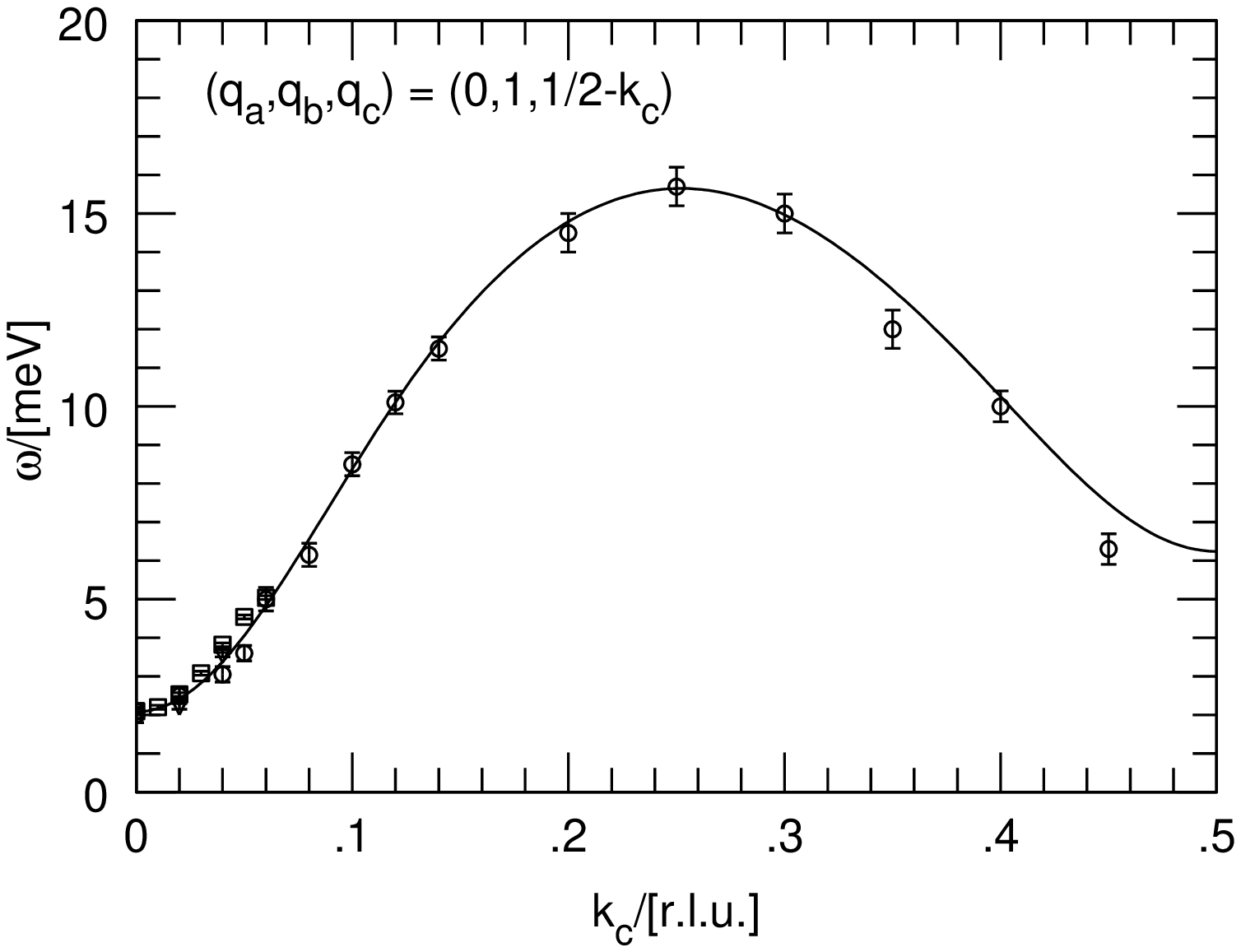,height=5.2cm,width=7cm}}
\end{picture}
\begin{picture}(8.2,5.4)(-1.55,0.4)
  {\psfig{file=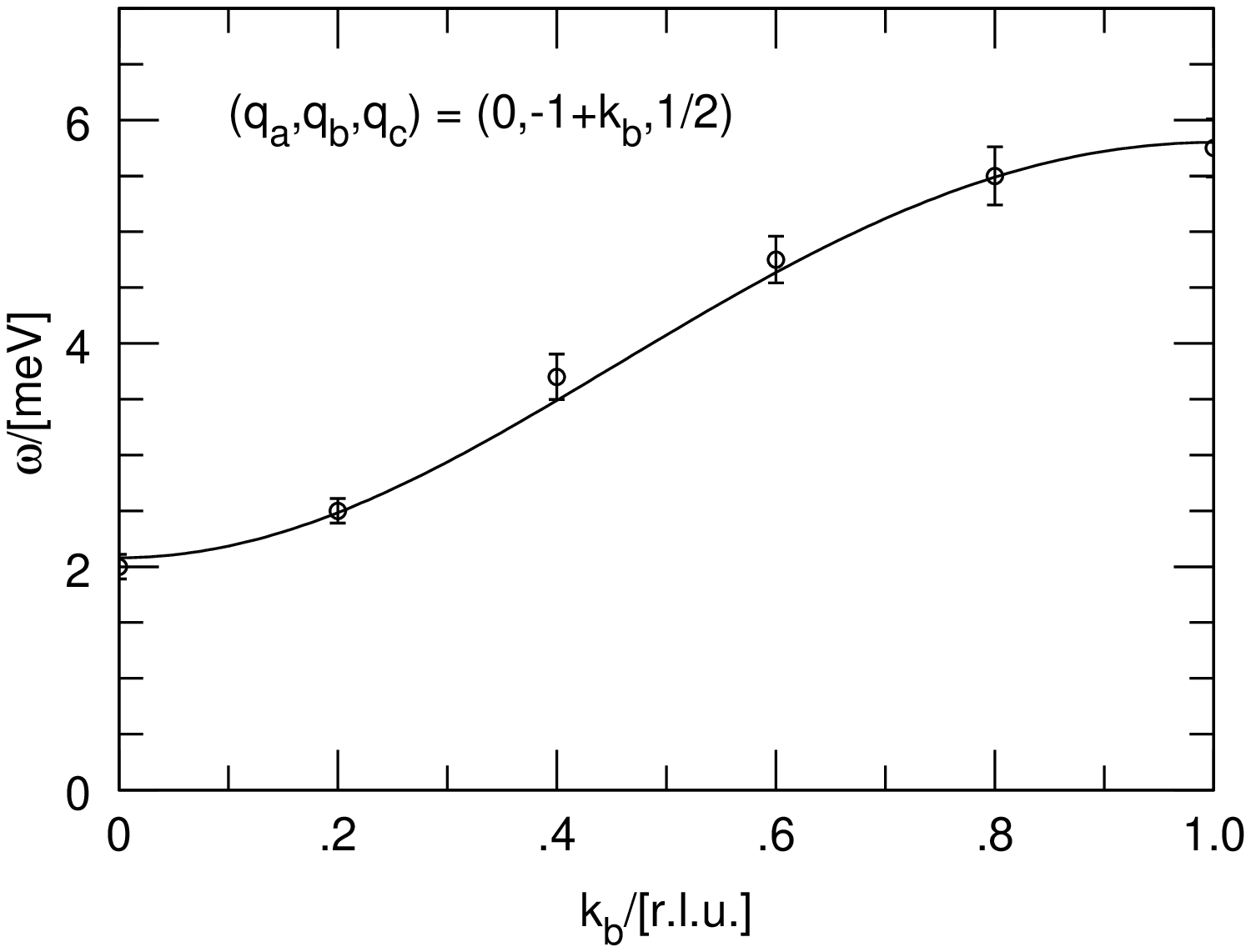,height=5.2cm,width=7cm}}
\end{picture}
\caption{
  a: Dispersion along the c-axis resulting from (\ref{result}) and
  (\ref{dispers}) at $(J,\lambda,\alpha,\mu,\beta) =
  (11.0{\protect\mbox{meV}},0.78,0,0.30,0.3)$ with experimental data
  after \protect\cite{regna96a}, fig.~8. Wave vectors in reciprocal
  lattice units (1 r.l.u. $=2\pi/$ lattice spacing).
  b: Same as in fig.~3a along the b-axis
  (after \protect\cite{regna96a}, fig.~5).}
\end{figure}
In fig.~3, a comparison between the theory and INS results is made. 
The best fit is obtained for $\alpha=0$ although recent
excellent $\chi(T)$-fits indicate that $\alpha=0.35$ for $T>30$ K is
realistic \cite{fabri97a,bouze97a}. The discrepancy can be due to the
truncation of the perturbation series at third order or to a change
of $\alpha$ across the spin-Peierls transition induced by the static
distortion or induced by a renormalization because of
non-adiabatic phonon effects. This remains to be clarified.

For the $\alpha=0$ fit the terms $\propto \lambda^2\mu^2, \lambda^4$
in (\ref{result}) are dropped.  The value $\lambda=0.78$ corresponding
to $\delta = 0.12$ is not as small as assumed previously ($\delta=0.03
\ldots 0.05$ \cite{riera95,casti95,haas95,regna96a}) but is of the
order of `first principles' values $\delta=0.21$ \cite{brade96a} or
$\delta\approx 0.1$ \cite{geert97}.
Note that the inclusion of higher order perturbation terms, enhancing
$\omega_{\rm\scriptstyle max}-\omega_{\rm\scriptstyle min}$ 
at constant $\lambda$, will lower the appropriate fit value for $\lambda$
even further.

The dispersion along the a-axis is accounted
for by adding to $\omega(\vec{q})$ as given by
(\ref{dispers}) a  term $\omega_a(\vec{q}) = 4t_a \cos(q_a)\cos(q_c)$
with $4t_a=-0.22$meV \cite{nishi94}.
This term is analogous to the third term in (\ref{basic}) since
the dimerization pattern alternates also in a-direction \cite{hirot94}.

The agreement is excellent in fig.~3a for $q_c<0.3$, where most data
points are available, and it is still very good for higher values.
The data points show clearly a small
shift to higher values of $q_c$ which results from the terms
proportional to $\cos(n q_c)$ with $n$ odd. Such terms are a signature
of interchain hopping  with alternating dimerization.
The agreement in fig.~3b is also good.

The fit in fig.~3 corroborates the approach to describe the 
dispersion as hopping on the lattice in fig.~2.
The parameters, however,
($J,\lambda, \alpha, \mu, \beta$) are not unambiguously fixed by the
dispersion data.
Good fits can also be obtained for smaller values
of $\mu$ and $\beta$, e.g.\ 
$(J,\lambda,\alpha,\mu,\beta) = (10.7\mbox{meV},0.78,0,0.14,0)$.
 This can be understood
by the fact that the main hopping effect along b is 
$\propto \mu(1-2\beta)$. 
To fix  the value of $\beta$ we deduced the coefficients 
$\chi_1\approx -7.8\pm 0.9 $meV and
$\chi_2\approx 26\pm 7$meV$^2$ in a large T-expansion of the 
magnetic susceptibility 
$\chi(T) = (\chi_0/T)(1+\chi_1/T+\chi_2/T^2 + ...)$
from the data in \cite{hase93a}.
Performing this expansion for the undimerized counterpart of the
lattice in fig.~1 yields
\begin{mathletters}
\begin{eqnarray}
\chi_1 &=& -(J_\lambda + J_\mu + J_\alpha +2J_\beta)/2 \\
\chi_2 &=& (J_\beta^2 +J_\lambda(J_\mu+J_\alpha) +
2J_\beta(J_\alpha+J_\mu+J_\lambda))/2
\end{eqnarray}
\end{mathletters}
where $J_\lambda =(1+\lambda)J/2$, $J_\alpha=J\lambda\alpha$,
$J_\mu=J\mu$, and $J_\beta=J\mu\beta$.
For parameters as in fig.~3,
one has $\chi_1=-7.5\mbox{meV}, \chi_2=30\mbox{meV}^2$
 while for $\beta=0$
one has $\chi_1=-5.4\mbox{meV}, \chi_2=6.9\mbox{meV}^2$.
This points to a certain amount of interchain frustration $\beta$
not too far away from $\beta\approx 0.5$.

Fig.~4 displays the overall form of $\omega(\vec{q})$ differing from
the schemes in \cite{loosd96a,bouch96}:
a) In fig.~4  $\omega(\vec{q})$ is roughly
constant for $q_c=0.25$, i.e.\ close to the maximum value of
$\omega(\vec{q})$, whereas it displays the same dependence on $q_b$
at all $q_c$ in \cite{loosd96a,bouch96}. b) According to 
\cite{loosd96a,bouch96} the smaller gap $E_{g1}\approx 2.1$meV is found at 
$\vec{q}=(0,1,1/2)$ and $(0,1,0)$; the larger  gap $E_{g2}\approx 5.5$meV
 is found at  $\vec{q}=(0,0,1/2)$ and $(0,0,0)$.
 In fig.~4, $E_{g1}$ and $E_{g2}$ are swapped at $q_c=0$.
This is very important in the interpretation of experimental
data (INS, Raman, ESR, FIR). The magnon DOS deduced from
fig.\ 4 is constant at the lower band edge (2d minimum),
it has a logarithmic singularity at $E_{g2}$ (2d saddle point),
but ressembles to a 1d, $1/\sqrt{\omega_{\rm\scriptstyle max}-\omega}$
DOS at the maximum due to the approximate degeneracy at $q_c = 0.25$.

Magnetic resonance data \cite{brill94,loosd96b} shows an excitation
from the ground state at  $E_{g2}$ which was  
interpreted to be the gap at $(0,0,0)$. 
But a vertical transition at $\vec{q}=(0,0,0)$ from $S=0$ (ground state)
to $S=1$ (excited states) 
is actually impossible due to total spin conservation.
\begin{figure}
\setlength{\unitlength}{1cm}
\begin{picture}(8.2,6)(0.8,0.65)
{\psfig{file=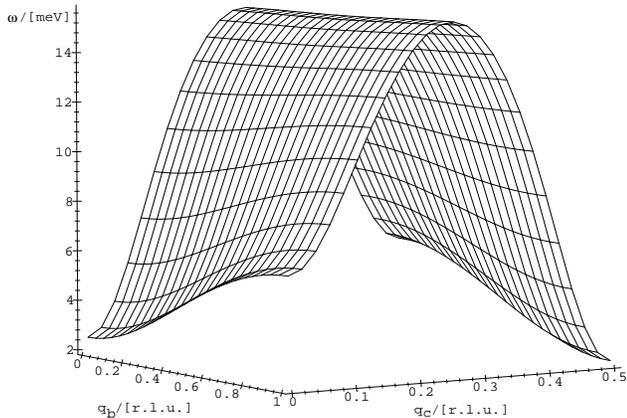,height=7.5cm,width=10cm,angle=270}}
\end{picture}
\caption{3D representation of the dispersion $\omega(\vec{q})$ for $q_a=0$
(parameters as in fig.~3).}
\end{figure}
For this reason and in view of 
fig.~4, a re-interpretation of this resonance is proposed.
In analogy to the magnetic resonances due to transitions
ground state $\to$ excited state in NENP and NINO 
(see e.g.\ \cite{brill95,sieli95}) the transition
 reflects the situation at $(0,1,0)$ \cite{bouch97}
because of the presence of  staggered magnetic fields due to
differently orientated $g$-tensors on neighboring chains in b-direction
\cite{pilaw97}. The Cu-atoms in each unit cell are inequivalent since
their crystal environments are differently orientated \cite{brade96a}
(cf.\ the situation for NENP/NINO \cite{sakai94bmitra94}).
This argument makes the magnetic resonance data
 compatible with fig.~4 (within 10\% accuracy as for NENP \cite{sieli95}).

Further INS investigation are desirable to make a more precise
analysis possible. It should be noted that the dispersion along
 the a-axis leads to {\em four}
different gaps in fig.~4. For the parameter set used in figs.~3 and
4, one has $\omega((0,1,1/2)) = 2.1$meV, $\omega((0,0,0)) = 2.5$meV,
$\omega((0,0,1/2)) = 5.8$meV, and $\omega((0,1,0)) = 6.2$meV.

The magnon dispersion described  will, at least qualitatively,
 also apply to the
currently investigated $\alpha'$-NaV$_2$O$_5$ where the 
dimerization is equally alternating
from chain to chain \cite{fujii96}. Such an alternation
appears to be generic for inorganic substances where 
bond angles are more easily changed than bond lengths.

In summary, we showed that the lattice distortion in the spin-Peierls
phase causes a dispersion which corresponds to hopping on a
lattice as in fig.~2. Thus the gap positions in the Brillouin zone are
revisited. A general perturbative result is given to enable
a quantitative analysis of the dispersion.
A larger dimerization ($\delta\approx 0.1$) in agreement with
orbital calculations is found.
This result covers also
chains and certain ladders which are a very active field as well 
\cite{dagot96}.
The magnetic resonance results are reinterpreted.

The author acknowledges fruitful discussions with J.P. Boucher, M. Braden,
W. Brenig, and W. Palme as well as  data communication  by L.~P. Regnault.
He thanks the LPS, Paris-Sud, for its hospitality.
This work was supported by an individual grant and by the SFB 341 of the DFG.


\begin{thebibliography}{10}
\bibitem{note1}
For larger values of $\delta$ one must distinguish
whether $\alpha$ is defined with respect to the average
or to the maximum value of the NN coupling.
\end{thebibliography}
\end{document}